\documentclass[%
 reprint,
 amsmath,amssymb,
 aps,
]{revtex4-1}
\usepackage{amsfonts,amssymb,amsmath}
\usepackage{graphicx}
\usepackage{dcolumn}
\usepackage{bm}
\usepackage[normalem]{ulem}
\usepackage{cancel} 

\usepackage{color}

\usepackage{geometry}
\usepackage{verbatim}
\usepackage{mathrsfs}
\usepackage{amsmath}
\usepackage{amssymb}
\usepackage{esint}

\usepackage{tikz}
\usetikzlibrary{shapes.misc}
\usetikzlibrary{decorations.markings}

\usepackage{soul}
\usepackage[unicode=true,pdfusetitle,
 bookmarks=true,bookmarksnumbered=false,bookmarksopen=false,
 breaklinks=false,
 colorlinks=true]
 {hyperref}

\begin{document}


\title{Unruh thermal hadronization and the cosmological constant}

\author{Antonia Micol Frassino} \email{frassino@fias.uni-frankfurt.de, Antonia.Frassino@etu.unige.ch}
\affiliation{D{\'e}partement de Physique Th{\'e}orique and Center for Astroparticle Physics,
Universit{\'e} de Gen{\`e}ve, 24 quai Ansermet, CH–1211 Gen{\`e}ve 4, Switzerland}

\author{Marcus Bleicher}%
\affiliation{Frankfurt Institute for Advanced Studies, Ruth-Moufang-Stra\ss e 1, D-60438 Frankfurt am Main, Germany}
\affiliation{Institut f\"ur Theoretische Physik,    Johann Wolfgang Goethe-Universit\"at Frankfurt am Main;}
\affiliation{GSI Helmholtzzentrum, Planckstrasse 1, 64291 Darmstadt, Germany}

\author{Robert B. Mann}\email{rbmann@uwaterloo.ca}
\affiliation{Department of Physics and Astronomy, University of Waterloo, Waterloo, Ontario, Canada}
\affiliation{Perimeter Institute for Theoretical Physics, Waterloo, Ontario, Canada}



\begin{abstract}
We use black holes with a negative cosmological constant to investigate aspects of the freeze-out temperature for hadron production in high energy heavy-ion collisions. The two black hole solutions present in the anti-de Sitter geometry have different mass and are compared to the data showing that the small black hole solution is in good agreement. This is a new feature in the literature since the small black hole in general relativity has  different thermodynamic behavior from that of the large black hole solution.  We find that the inclusion of the cosmological constant (which can be interpreted as the plasma pressure) leads to a lowering of the temperature of the freeze-out curve as a function of the baryochemical potential, improving the description previously suggested by  Castorina, Kharzeev, and Satz.
\begin{description}

\item[PACS numbers]
\end{description}
\end{abstract}

\pacs{Valid PACS appear here}
\maketitle


\section{\label{sec:Introduction}Introduction}

In recent years relativistic heavy-ion collisions at high energies have become a laboratory for exploring new states of matter \cite{Adcox:2004mh,Back:2004je,Adams:2005dq,Aad:2013xma} and for testing exciting new ideas for the description of these novel states \cite{McLerran:2008ua,Gelis:2009wh,Song:2010mg,Heinz:2015lpa,Mrowczynski:2016etf}.  It is currently understood that in these reactions a strongly interacting quark-gluon plasma (QGP) forms and behaves like a nearly perfect liquid, i.e. the shear viscosity is very small \cite{Gyulassy:2004zy,Heinz:2015lpa}. However, the approach to apparent local equilibrium is still under debate \cite{Strickland:2017kux,Romatschke:2017acs}. Current estimates of these effects are based on hydrodynamical modelling (mostly at vanishing baryochemical potential $\mu_B$) \cite{Ryu:2015vwa,Gale:2012rq,Heinz:2013th,Karpenko:2015xea}, lattice QCD \cite{Cheng:2008zh,Borsanyi:2013bia}  or, via  gauge/gravity duality, 
on strongly coupled dual non-Abelian plasmas with a large number of colors \cite{Kovtun:2004de,Buchel:2003tz}. 
 
Lattice QCD is the main non-perturbative technique that can be used to study strongly interacting QCD physics.  However,  when $\mu_{B} \neq 0$, lattice approaches are affected by 
the sign problem of the fermion determinant. Various alternative tools have been developed to address this issue, allowing one to investigate, at least in principle, small chemical potentials on the lattice (see for instance \cite{Fodor:2001au,Gavai:2003mf,Allton:2003vx,Philipsen:2012nu}).  It is therefore  important to have other theoretical ways to study QCD phenomena in the strongly coupled regime at nonzero $T$ and $\mu_{B}$.

Gauge/gravity duality has provided  relevant insights into the study of real-time non-equilibrium dynamical phenomena (for a review see \cite{CasalderreySolana:2011us} and references therein). |
The duality allows one to calculate physical quantities describing a strongly coupled gauge theory on the boundary of a $d+1$ space with a gravitational theory.
Holographic descriptions of strongly interacting systems (bottom-up models) cannot be obtained in general from top-down string theory constructions and are based on the conjectured validity of the duality under more general circumstances.

Here, however, we consider the analogy between gravity and gauge theories from another point of view, based on a conceptual framework proposed some time ago  \cite{NCimento76,PhysRevD.3.867, PhysRevD.18.4596} and  developed in more detail over recent years \cite{Castorina:2007eb,Becattini:2008tx}. The analogy relies on the confinement property of QCD, i.e. the fact that QCD forbids colored constituents to exist in the physical vacuum. It resembles in some way the phenomenon of gravitational confinement of matter inside a black hole.
Indeed, a black hole can be regarded as a solution to Einstein's equations defined by a confining potential. The fate of matter near a black hole (within its innermost stable circular orbit) is to inevitably  fall  through the event horizon in the absence of countervailing forces.
 The application of the quantum mechanics to black holes resulted in the discovery of their thermal emission \cite{1975CMaPh43199H}. Soon after Hawking's pioneering work, Unruh showed that an observer under uniform acceleration $a$ experiences a  thermal bath at temperature $T=a \hbar / 2 \pi$ \cite{PhysRevD.14.870}.

In other words confining potentials in general lead (quantum mechanically) to an intrinsic temperature. This led to the proposal that  quarks in a confining potential are also associated with an effective temperature for hadrons \cite{Grillo:1979wu}. 
More specifically, following the connection between gravitational properties and particle physics, a conjecture was put forward that color confinement causes  the physical vacuum to form an event horizon for quarks and gluons that can only be crossed  by quantum tunneling \cite{Castorina:2007eb,Kharzeev:2005iz}.
In this sense hadron production corresponds to a form of Hawking-Unruh radiation in QCD.

This analogy is also supported by two additional facts in black hole physics:
\begin{itemize}
    \item[(i)]  The metric of a system in uniform acceleration, the Rindler metric, is equivalent to the near-horizon approximation of the black hole metric if the acceleration is equal to the surface gravity $\kappa$;
    \item[(ii)]  Hawking radiation is a quantum phenomenon associated with pair-creation near the event horizon and the tunnelling of particles 
    \cite{Parikh:1999mf,Kerner:2006vu,Vanzo:2011wq}, in analogy with  string breaking \cite{Casher:1978wy,Andersson:1983ia,Becattini:1995if} and pair creation in systems with uniform acceleration. 
\end{itemize}
Within the context of the above conjecture,  one may consequently propose the following hypotheses: (a) the hadronic freeze-out temperature at high energy is an Unruh temperature; (b) the associated Rindler horizon can be identified with a ``color blind'' horizon dynamically produced by the color charge confinement during the $q \bar{q}$ production.

Previous approaches toward a concrete realization of this conjecture have been
 restricted to a charged black hole in an asymptotically flat background \cite{Castorina:2016xrm}. However (as well shall see) the freeze-out temperature obtained by this analogy does not describe the data at low $T$ and large $\mu_B$ \cite{Castorina:2016xrm}. 
 
In this paper we enlarge the analogy presented in \cite{Castorina:2007eb,Castorina:2014fna} to a charged black hole
in anti-de Sitter spacetime (AdS). We vary the AdS curvature radius and see if it has a particular counterpart in the description of the freeze-out. 
The thermodynamics of an AdS black hole has features that we will see are important in the analogy, namely a minimum temperature $T_{\text{min}}$, which occurs when the horizon radius is of the order of the characteristic radius of the AdS space \cite{hawking1982}.  Above $T_\text{min}$ there are two possible black hole solutions with different radii.
We find that the freeze-out temperature is well described by the Hawking temperature of the small AdS charged black hole, providing a very favourable fit to current data.

The sections are organized as follows: in Sec. \ref{sec:gravP} we review  anti-de Sitter spacetime and the hypothesis of interpreting a variable negative cosmological constant as pressure. In sec. \ref{sec:QCD-BH} we review the connection proposed in \cite{Castorina:2007eb,Castorina:2014fna} between black holes and QCD. Sec. \ref{sec:AdSQ} presents to our calculations and results. Finally, sec. \ref{sec:conclusions} is dedicated to the conclusions.

\section{Gravitational pressure}\label{sec:gravP}

The study of  black hole thermodynamics in the presence of a negative cosmological constant $\Lambda$ (i.e. an AdS background) has exhibited very interesting properties  \cite{Hawking:1982dh}, subsequently opening the way to further insights into string theory \cite{Witten:1998zw} and thermodynamic phase transitions \cite{Chamblin:1999tk,Chamblin:1999hg,Kubiznak:2012wp}.
The notion that the cosmological constant itself might be considered as a dynamical variable was initially suggested by Teitelboim and Brown in \cite{Teitelboim:1985dp, Brown:1988kg}, while the relative thermodynamic term was included only later into the first law of black hole thermodynamics \cite{Creighton:1995au}.
The idea of associating the cosmological constant with pressure was then explored in different ways  \cite{Caldarelli:1999xj,Padmanabhan:2002sha,Kastor:2009wy}. 
In contrast to an asymptotically flat Schwarzschild black hole, a black hole in an AdS background with sufficiently large radius (as compared to the AdS radius $\ell$) has positive specific heat and so can be in stable equilibrium at a fixed temperature (where the AdS space mimics a gravitational box). 
Depending on the temperature, it can also be subjected to a phase transition to pure radiation known as the Hawking-Page transition  \cite{Hawking1983}. 
In the framework of the AdS/CFT duality, this transition was later associated with a confinement/deconfinement phase transition in the field theory on the boundary \cite{Witten:1998qj}.
The idea of associating the cosmological constant $\Lambda$ and hence the AdS radius $\ell$, see Eq. \eqref{eq:Ptol}, with a pressure (along with the notion of a conjugate thermodynamic volume) requires the generalization of the
laws of black hole mechanics \cite{Kastor:2009wy}. The pressure can be defined as
\begin{equation}
    P=- \frac{\Lambda}{8 \pi G} = \frac{\left( d-1 \right) \left( d-2 \right) }{16 \pi G \ell^2}, \label{eq:Ptol}
\end{equation}
where $d$ is the number of spacetime dimensions and $G$ is the Newton constant.
The resultant generalized first law of black hole thermodynamics is
\begin{equation}
    \delta M = T \delta S + V \delta P + \Omega \delta J + \Phi \delta Q \label{eq:1stlaw}
\end{equation}
where $J$ is the black hole angular momentum and $Q$ the charge.
The quantity $M$ is the conserved charge associated
with the time-translation Killing vector of the spacetime. The entropy is related to area of the black hole event horizon according to $S= A / \left(4 \hbar G\right)$  
and the Hawking temperature is $T = \hbar \kappa/ \left( 2 \pi \right)$ where $\kappa$, as mentioned before, is its surface gravity. The conjugate thermodynamic volume to the pressure is defined as $V \equiv \left( \partial M / \partial P \right)_{S,Q,J}$. From this viewpoint the confinement/deconfinement phase transition can be understood as a solid/liquid phase transition  \cite{Kubiznak:2014zwa,Kubiznak:2016qmn}.

Because of the presence of the cosmological constant in Eq. (\ref{eq:1stlaw}), the mass
$M$ cannot be interpreted as usual as the internal energy of the system. Rather, $M$ can be understood as the gravitational version of the chemical enthalpy \cite{Kastor:2009wy}, i.e., the total energy of a system containing both the energy $P V$ needed to displace the
vacuum energy of its environment and its internal energy $E$ \cite{Kubiznak:2014zwa}.

The definition of the cosmological constant as pressure will allow us to proceed with the study of the freeze-out temperature using the physical parameters $M,Q,P$ of the extended phase space. Using the pressure in this way we introduce the corresponding thermodynamic volume.
Therefore, for each value of $P$ we will consider the corresponding volume of the (regularized) spacetime at a fixed time slice that is the volume inside the black hole.

\section{QCD and Black Holes} \label{sec:QCD-BH}

Classically, a black hole just absorbs matter. On the quantum level, however, 
 matter inside the black hole (i.e., its constituents: hadrons, leptons and photons) has a non-vanishing probability to escape by tunnelling through the barrier of the event horizon.  The transmitted radiation is thermal and ensures color neutrality. The thermal behavior of black holes is fully encoded in the thermodynamic description:  Hawking radiation cannot give any information related to the internal state of the black hole. 

In all collisions $e^{+} e^{-}$, $pp$, $p \bar{p}$, $\pi p$, etc..., including nucleus-nucleus scattering, particle production likewise exhibits thermal behaviour that seems to occurs at the same temperature \cite{Becattini:1995if,Becattini:1997rv,Cleymans:1999st,BraunMunzinger:2003zd,Andronic:2005yp,Becattini:2005xt,Andronic:2008ev} (see also \cite{Muller:2017vnp}).  This feature motivated the proposal \cite{Castorina:2007eb}   that in relativistic heavy-ion collisions at large $\sqrt{s}$, corresponding to zero baryochemical potential \cite{Becattini:2005xt}, the hadronic freeze-out temperature $T$ 
is an Unruh temperature. Hadronization can be seen as the QCD counterpart of   Hawking radiation. Relativistic heavy-ion collisions are expected to create a quark-gluon plasma, QGP, that  goes through an expansion at close to the speed of light,  emitting   radiation and then undergoing a deconfinement-confinement phase transition to a hadron gas. Experiments at RHIC and LHC indicate that a thermalized QGP is formed in collisions between two heavy nuclei at center-of-mass energies of the order $100$ GeV per nucleon \cite{Ackermann:2000tr,Aamodt:2010pb,yagi2005quark}.
Particle spectra in the expanding geometry (i.e., in the Bjorken or Milne coordinates) look analogous to those in Rindler coordinates. In this sense  an event horizon is associated with hadron production in a similar spirit as horizon formation in analog gravity models, see e.g. \cite{Visser:1997ux}.

In this picture hadrons produced in the collisions are assumed to be formed in equilibrium and to follow a thermal distribution as consequence of the random distribution of quarks and antiquarks, which are entangled in such a vacuum.  

\subsection{Charged Black Hole}

From now on we set $c=1$ and write the Einstein-Maxwell action for a black hole charged under a Maxwell
field $F=dA$ as follow:
\begin{equation}
    S_{EM}\left[ g,A \right]=\frac{1}{16 \pi G} \int d^4 x \sqrt{|g|} \left[ R- 4 \pi G F^2 \right] 
\end{equation}
where $R$ is the Ricci scalar and $g$ the determinant of the metric. In these units,  $g_{\mu \nu}$ is dimensionless and the electric charge is defined:
\begin{equation}
    Q=\frac{1}{8 \pi} \int_{S^{2}_{\infty}} \star F,\label{eq:charge}
\end{equation}
Since the Maxwell equations are satisfied if the Einstein equations are, one has only to solve the latter with the trace subtracted
\begin{equation}
    R_{\mu \nu} = 8 \pi G \left[ F^{\;\; \rho}_{\mu}  F_{\nu \rho} - \frac{1}{4} g_{\mu \nu} F^2 \right]
\end{equation}
plus the Bianchi identity. The static, spherically symmetric solution is the Reissner-Nordstr\"om (RN) solution  
\begin{eqnarray}
ds^2 &=& f \left( r \right) dt^2 - f^{-1}\left( r \right) dr^2 -r^2 d\Omega^2 \label{sssmet}\\
F_{tr} &=& \frac{Q}{r^2 }\\ 
f \left( r \right) &=& \frac{\left( r- r_{+} \right)\left( r-r_{-} \right)}{r^2}
\end{eqnarray}
where
$r_{\pm}=GM\pm \sqrt{\left(G^2 M^2 - 4\;\pi G Q^2\right)}$ are the inner and outer horizons, $Q$ is the electric charge normalized as in (\ref{eq:charge}) and $M$ is the ADM mass.
\\ 
The first law (\ref{eq:1stlaw}), with angular momentum $J=0$, takes into account the possible changes in the black hole mass due to changes in the charge. The temperature can be written as \cite{Ortín201501}
\begin{equation} \label{Tcharged}
    T(M,Q) = T(M,0) \frac{4 \sqrt{1- \frac{4 \pi Q^2}{G M^2} }}{\left( 1 + \sqrt{1-\frac{4 \pi Q^2}{G M^2}} \right)^2}
\end{equation}
and, substituting in the previous equation (\ref{Tcharged}) the value for the mass obtained from the horizon equation $f\left( r \right) = 0$, this can be expressed in the simple form
\begin{equation}
     T(M,Q) = T(M,0) \left(1 - G^2 \Phi^4 \right)\label{eq:TGPhi}
\end{equation}
where $\Phi \left( r_{+}\right) = q / r_{+}$
is the electrostatic potential on the horizon, with  $q=4 \pi Q$. 

The  temperature $T(M,Q)$ satisfies the first law of black hole thermodynamics \cite{bardeen1973} 
\begin{equation}
    \delta M = T \delta S + \Phi \delta Q \label{eq:1lawH}
\end{equation}
where the entropy is $S= \pi r^{2}_{+}/G$.  
The Smarr formula (an integral version of (\ref{eq:1lawH})) takes the form $M=2TS +Q\Phi$.

The conjectured equivalence between gravitational confinement and color confinement 
maps the electric potential and the gravitational constants to the baryochemical potential $\mu$ and the string tension $\sigma$.  It
takes the form \cite{Tawfik:2015fda,Castorina:2016xrm}
\begin{equation}
    \left\{ \Phi, G \right\} \rightarrow  \left\{ \mu, \frac{1}{2 \sigma} \right\} \label{eq:equiv}
\end{equation}
and the freeze-out parameters $T_f$ and $\mu_B$ (that we will simply call $T$ and $\mu$) can be calculated from (\ref{eq:TGPhi}). 
Using this equivalence and introducing a new constant $\bar{\mu}$ function of the string tension,   eq. (\ref{eq:TGPhi}) can be expressed as
\begin{equation}
    T \left( \mu, \sigma \right) = T_0 
    \left[ 1 - \left( \frac{\mu^{2}}{\bar{\mu}^{2}}  \right)^2 \right], \label{eq:Tcharged}
\end{equation}
 where $T_0$ is the temperature of the black hole when the charge is zero.
Note that in previous calculations \cite{Castorina:2016xrm}, the Unruh mechanism could \emph{a priori} describe the freeze-out process only as long as  $\mu \ll m_{\rm proton}$. 
Also keep in mind that the comparison with a black hole can provide an explanation only for the production of new hadrons in high-energy collisions, while it does not describe the contribution of the nucleons that can be already present in the initial state of heavy-ion collisions  \cite{Castorina:2014fna}.

However, assuming the relation (\ref{eq:equiv}) between the electrostatic potential on the horizon and the baryochemical potential \cite{Castorina:2008gf}, then the freeze-out parameters $T$ and $\mu$ can be calculated and compared with the freeze-out parameters inferred from other models and from data. 

In particular, in our analysis, we fix (i) the value of the baryochemical potential at $T=0$ to the value of the proton mass   $\mu = \bar{\mu}=0.938$ GeV (and we also re-scale the result from ref. \cite{Castorina:2016xrm} to this value) and (ii) $T_0$  to the lattice result $T_L=0.155$ GeV.
The resulting freeze-out temperature for the case of a charged black hole in asymptotically flat spacetime \cite{Castorina:2016xrm} corresponds to the dotted line in Fig. \ref{fig:figure1}.

\section{Charged Black Hole in AdS}\label{sec:AdSQ}

The original analogy between color confinement and gravitational confinement was proposed in the context of  a Schwarzschild black hole \cite{Castorina:2007eb}. However, this solution has uncommon thermodynamic properties such as negative specific heat; it also does not exhibit a Hagedorn temperature. The analogy was later extended to a charged black hole \cite{Castorina:2016xrm}.  Here we consider a further extension to a charged black hole in an anti-de Sitter background.

In Einstein-Maxwell-anti-de-Sitter theory the action can be written as
\begin{equation}
    I=\frac{1}{16 \pi G} \int d^{4} x \sqrt{|g|} \left[ R - 4 \;\pi \;G F^2 + \frac{6}{\ell^2} \right]
\end{equation}
with $\Lambda= - 3/\ell^2$, where $\ell$ is the characteristic AdS length.
The solution to the Einstein-Maxwell-AdS equations corresponding
to a $d=4$ dimensional charged-AdS black hole is given by \eqref{sssmet} but now with
\begin{equation}
    f \left( r \right)=1-\frac{2GM}{r}+\frac{4 \pi G Q^2}{r^{2}}+\frac{r^2}{\ell^2}
\end{equation}
The temperature $T = f^{\prime}\left( r_{+} \right)/\left( 4 \pi \right)$  is
\begin{equation} \label{eq:TAdS}
    T \left( r_{+}, Q, P \right) = 
    \frac{1}{4 \pi r_{+}}\left[ 1 + 8\pi G P r^2_{+} -  \frac{4 \pi G Q^2}{r^2_{+}} \right],
\end{equation}
where $r = r_{+}$, determined from $f(r_{+}) = 0$, is the radius for the outer (event) horizon. 
The other thermodynamic quantities are 
\begin{eqnarray}
S &= & \frac{A}{4 G}= \frac{\pi r^{2}_{+}}{G},\\
V &= & \left( \frac{\partial M}{\partial P} \right)_{S,Q} = \frac{4}{3} \pi r^{3}_{+},\\
\Phi &= & \left( \frac{\partial M}{\partial Q} \right)_{S,P}= 4 \pi \frac{Q}{r_{+}},
\end{eqnarray}
and the Smarr relation is $M=2TS - 2PV + \Phi Q$.

In the case of charged AdS black hole, it is not possible to write a simple expression for the temperature as function of the physical parameters, like the mass $M$ and the electrostatic potential $\Phi$, as in Eq. (\ref{eq:TGPhi}).
However, it is possible to evaluate the temperature as a function of $\Phi$ numerically once the couple $(M,G)$ are fixed while the parameter $P$ can vary in a certain range.
Fixing the value of the temperature $\tilde{T}_0$ for the uncharged AdS black hole  and the value of the charge $\tilde{Q}$ to the value that gives zero temperature 
\begin{eqnarray}
T (M,P, Q=0) &\equiv& \tilde{T}_{0},\label{eq:esremoAdsT0}\\
T(M,P,\tilde{Q})&=&0,\label{eq:esremoAdsmubar}
\end{eqnarray}
allow us to calculate the couple $(M,G)$. 
In contrast to the case of a charged black hole in a flat background presented in the previous section, the space of the parameters here is constrained.
\\
\paragraph{Constraints from eq. \eqref{eq:esremoAdsT0}.}
An uncharged black hole in an AdS background has a minimum temperature $T_{\min}$ such that for $T<T_{\min}$ there is only thermal AdS.
At fixed $\tilde{T}_0$ this implies that not all  values of $P$ are allowed.
The temperature for an uncharged AdS  black hole as function of the horizon radius is
\begin{equation}
    T \left( r_{+},  P, Q=0 \right) = 
    \frac{1}{4 \pi r_{+}}\left[ 1 + 8\pi G P r^2_{+}  \right], \label{eq:TunCharg}
\end{equation}
and the requirement $T \left( r_{+},  P, Q=0 \right) \geq T_{\min}$ 
gives the constraint
\begin{equation}
\tilde{T}_{0} \geq \sqrt{2 G P / \pi}.\label{eq:constrP}
\end{equation}
Once $G$ is fixed, the inequality \eqref{eq:constrP} bounds the allowed values of the pressure.

In an AdS background, above the minimal value of the temperature $T>T_{\min}$ there are always two black hole solutions: a \emph{large} black hole and a \emph{small} black hole (depending on the value of the mass).
Note, also, that the pressure $P = -\Lambda/8\pi G$ as defined in Eq. \eqref{eq:Ptol} is a combination of both the Newton gravitational constant and the AdS radius.
\\
\paragraph{Constraints from eq.  \eqref{eq:esremoAdsmubar}.} In the case of the extremal black hole (i.e., when the black hole temperature is zero while the other parameters are nonzero ($P,Q,M,G \neq 0$)),
it is possible to find a formula for the mass as a particular combination of the other parameters 
\begin{equation}
    M=\frac{\sqrt{G \Phi ^2-1} \left(2 G \Phi ^2+1\right)}{6 \sqrt{2 \pi } G^{3/2} \sqrt{P}}
\end{equation}
yielding another condition:  $\Phi< G^{-\frac{1}{2}}$. \\
\begin{figure}
\vspace{-0.5cm}
\includegraphics[scale=0.45]{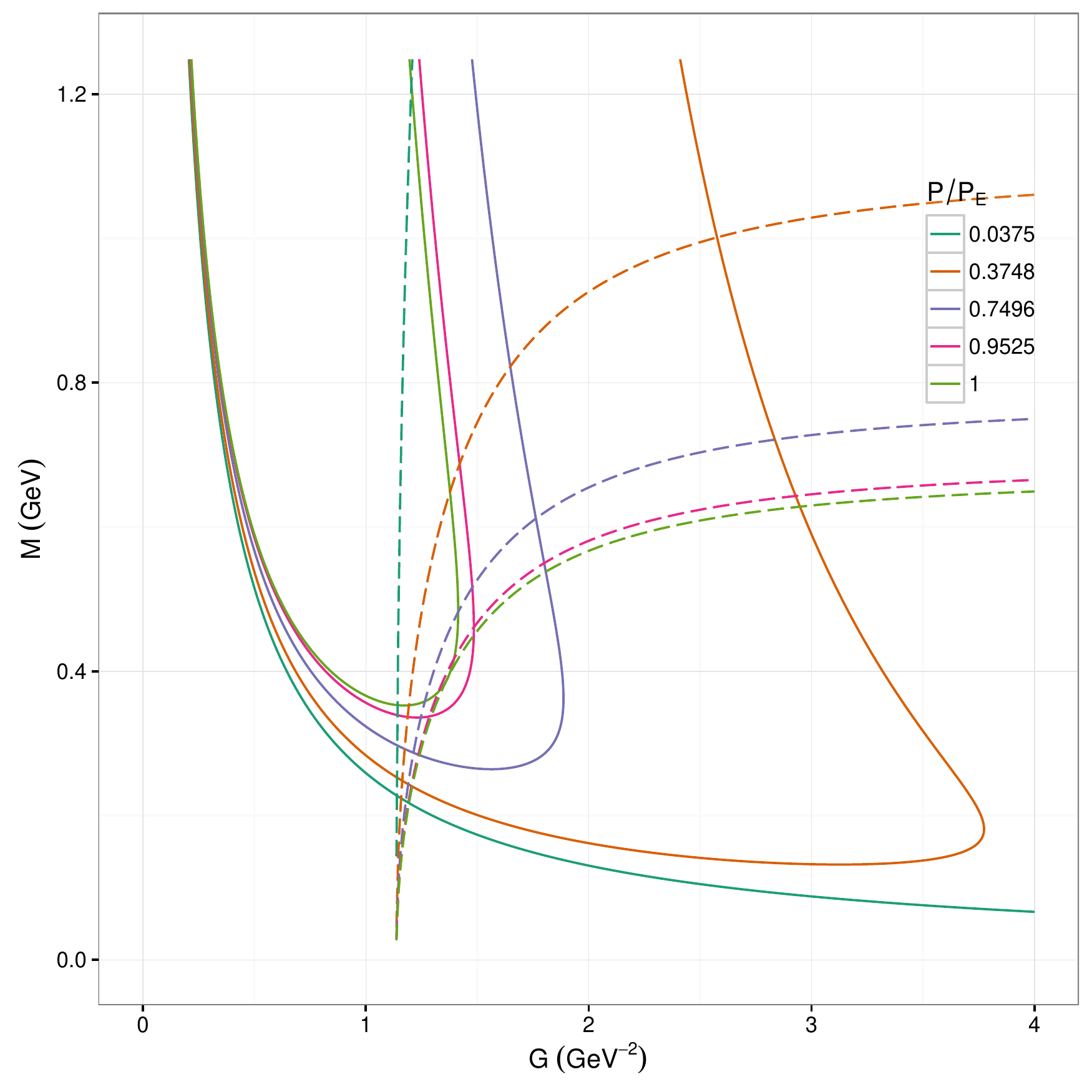}
\caption{The curves denote the parameter space where the conditions \eqref{eq:esremoAdsmubar} (dashed) and  \eqref{eq:esremoAdsT0} (solid)  are satisfied.
The temperature for the uncharged AdS black hole $T_0$ in \eqref{eq:esremoAdsT0} is set to be $T_L$. For $P<P_E$,  
each dashed curve 
intersects with each solid curve at two different points. We  denote these points 
as ``\emph{small}'' and ``\emph{large}'' black hole solutions, corresponding to the  value of their mass.  At $P=P_E$ there exist only one pair of values $(G,M)$ (the dashed and solid green curves meet at one point).   \label{fig:figure0}}
\par
\end{figure}
\begin{figure*}[t!]
\vspace{-0.5cm}
\includegraphics[scale=0.65]{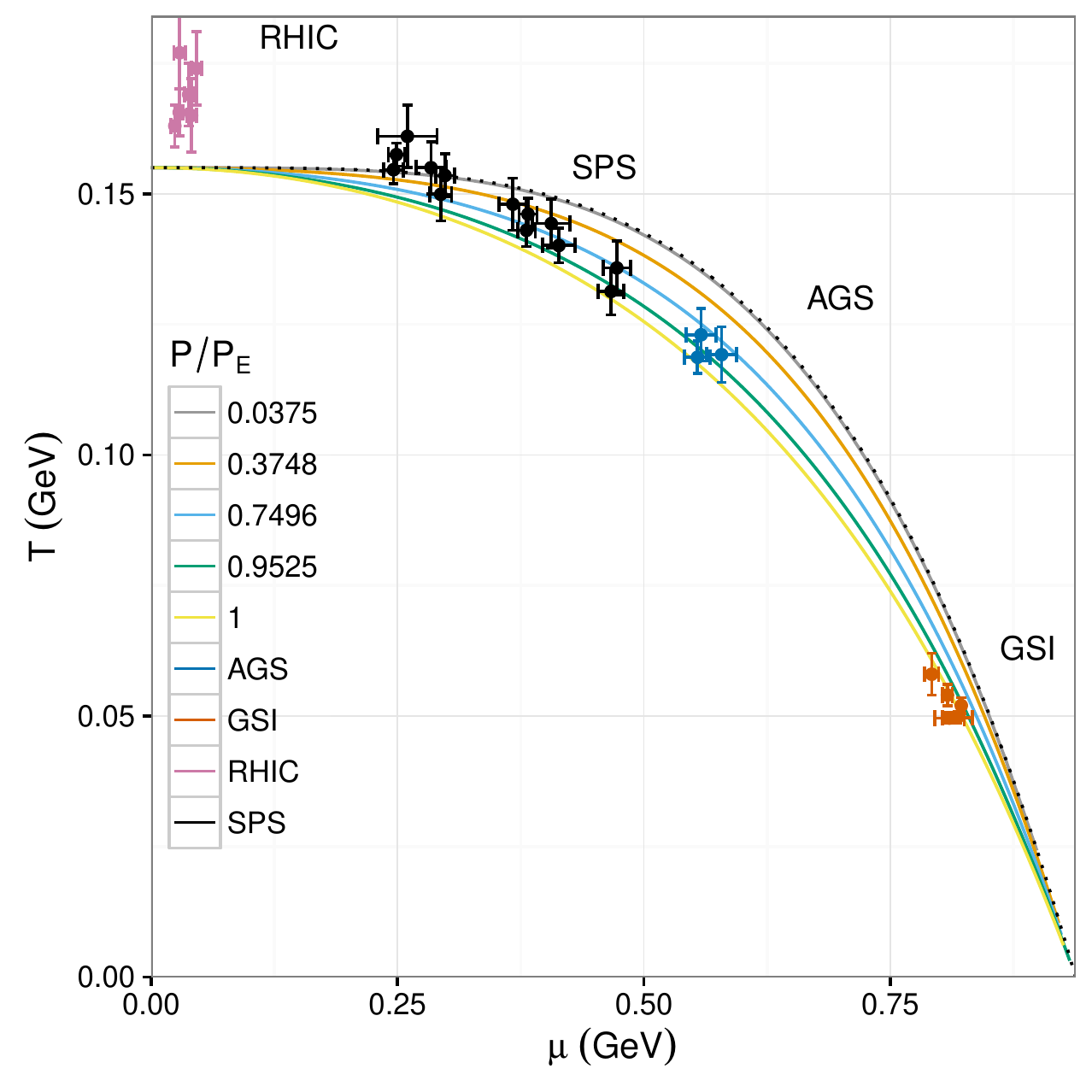}
\includegraphics[scale=0.65]{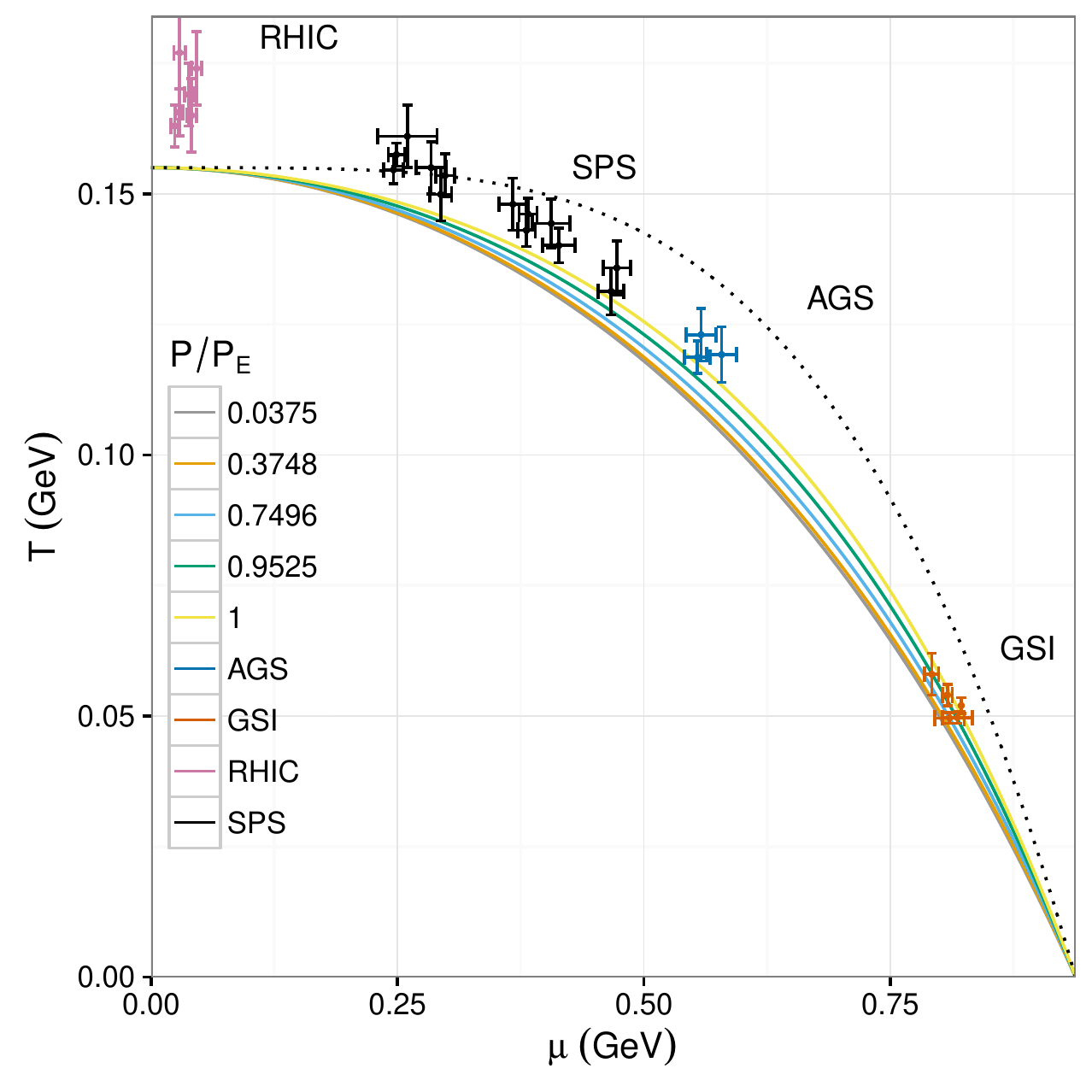}
\caption{Freeze-out data in the statistical hadronization model reported in \cite{Cleymans:2005xv} compared with the criteria discussed in the text for the \emph{small AdS black hole solution} (left panel) and for the \emph{large AdS black hole solution} (right panel).
The dashed curve is for the case $P=0$ \cite{Castorina:2016xrm}. The other lines are the temperatures for different values of $P$ (and corresponding values of $(G,M)$).  \label{fig:figure1}}

\par

\end{figure*}
\paragraph{QCD parameters}
Fig. \ref{fig:figure0} depicts the curves where \eqref{eq:esremoAdsT0} and \eqref{eq:esremoAdsmubar} are satisfied in the plane $(G,M)$ for different values of the pressure. The intersection points are the values where both of the equations are satisfied;  they can be obtained numerically.

As in the previous section,  $\tilde{T}_{0} = T (M,P, Q=0)$ is the temperature when the baryochemical potential is zero (i.e., an uncharged AdS black hole) and can be fixed to be the lattice temperature $T_L$.   Eq. \eqref{eq:esremoAdsmubar}, instead, is the condition that defines an extremal black hole (i.e. a black hole with zero temperature), obtained when the chemical potential is of the order of the proton mass $\bar \mu=0.938$ GeV. This choice is motivated by the observed universal freeze-out curve $\langle E\rangle / \langle N\rangle \approx 0.94-1$GeV \cite{Cleymans:1998fq} and the fact that a nucleus dissolves at rather low excitation energies, setting the decoupling condition at $T\approx 0$ to a value near $\mu_B\approx m_{\rm proton}$.

As noted above, and in contrast  to what happens for an asymptotically flat charged black hole, the concurrent fulfillment of the conditions  \eqref{eq:esremoAdsT0} and \eqref{eq:esremoAdsmubar} at a fixed pressure  $P$ provides two solutions, one, or no solutions for the couple $(M,G)$. We maintain the nomenclature used for the uncharged AdS black hole, denoting a large/small  black hole  as being one
with large/small mass (even though the solutions have different values of $G$). In particular, there exists an extreme pressure $P_E$ where the two solutions collapse to one, bounding the allowed pressures $ 0 \le P \le P_E$.

These constraints are given by the AdS geometry and form an  important point of our analysis: unexpectedly the intersections of the two constraint curves yield parameters in the range of the experimental data.

\section{Analysis}

In Fig. \ref{fig:figure1} we plot the freeze-out temperature as function of $\mu$ using the analogy (\ref{eq:equiv}) for the \emph{small} mass (left) and the \emph{large}  mass (right) solutions  compared to the data \cite{Cleymans:2005xv}. Each curve corresponds to a different value of the pressure in the range $0< P/P_E \leq 1$.  We see that the introduction of a pressure $P\neq 0$ moves the temperature curves below the curve at $P=0$. The value of $T_L$ fixes the couple $(M,G)$ defining, therefore, the \emph{small} and \emph{large} black holes.

We see from the right-hand side of figure \ref{fig:figure1}  that the large black hole solution is not continuously connected to the curve at $P=0$ (the dashed line, inserted for reference). This is because the large \emph{uncharged} black hole solution does not exist when $P=0$. The large black hole temperature is, instead, always bounded by the extremal pressure $P_E$. For a fixed G   we could say that the large black hole is colder than the corresponding small black hole at the same value of the pressure. 

However from the left-hand side of figure \ref{fig:figure1}, we see that the temperatures for the small black hole solutions are all bounded by the curve at extremal pressure $P_E$ (yellow line) and the curve at $P=0$ (dashed line).  The data are commensurate with 
$0.95 < P/P_E < 0.75$. 
 
The parameters of the gravitational analog used in this work are fixed to reproduce two crucial relations: (i) the baryochemical potential at zero temperature is given by the proton mass and, (ii) the temperature $T$ at zero baryochemical potential $\mu=0$ 
coincides with the corresponding lattice QCD result (which is the same constraint that the Einstein-dilaton equations of the holographic dual satisfy). 

Indeed, in the holographic gravitational models, there is a $U(1)$ charge to 
mimic the baryon charge and the corresponding chemical potential $\mu$ while a real scalar field in the bulk is used to break the conformal invariance  and therefore to take into account the effects from the QCD running coupling. 
In particular, a realistic description of a nonconformal QGP is given by a dilatonic gravity dual \cite{Gubser:2008ny} 
where there is a scalar field $\phi$ that is coupled to the metric $g_{\mu \nu}$ and that is responsible for breaking the conformal symmetry of the theory in the IR regime, 
reproducing the effects of a $\Lambda_{QCD}$ generated dynamically (for some studies see \cite{Critelli:2017oub,Rougemont:2015wca}).

The map presented in \eqref{eq:equiv} suggests that there could also be a corresponding quantity associated with the cosmological constant that can be varied. In particular, once the pressure is fixed to the best value that is in accord with the data (see Fig. \ref{fig:figure1})   the corresponding volume is also fixed, giving the radius of the black hole that is in agreement with the data.
Since the QCD plasma pressure is isotropic within a 25\% approximation during the relevant time scales \cite{Zhang:2008kj}, this suggests that $\Lambda$ can be interpreted as the pressure, fitting  with  the $PV$ interpretation in black hole thermodynamics.

\begin{figure}
\vspace{-0.5cm}
\includegraphics[scale=0.25]{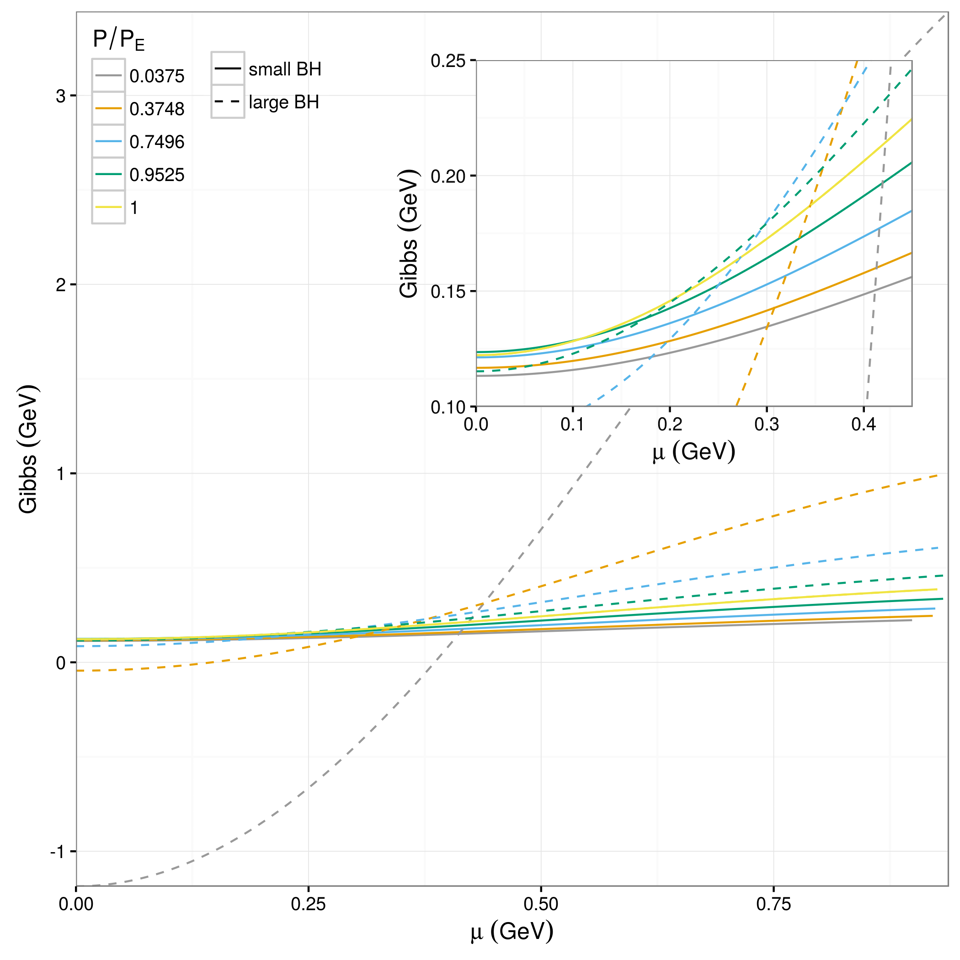}
\caption{A plot of the Gibbs Free energy with respect to small and large solutions given by the various
curves in figure 1.  We see that   the small black hole has a larger  Gibbs free energy
for small chemical potential.  \label{fig:figure3}
}
\end{figure}

That the small black hole provides a better fit to the data -- the decay of an unstable QCD plasma -- 
is consistent with it being
the thermodynamically unstable gravitational configuration. In  Fig. \ref{fig:figure3} we depict a comparison of the Gibbs Free energy with respect to the two possible solutions given by the various
curves in Fig. \ref{fig:figure1}. This  indicates that   for small
chemical potential, the small black hole has a greater Gibbs free energy
than the large black hole.  As the chemical
potential increases, the Gibbs free energy of the large solution grows faster than
that of the small one, and eventually the small solution becomes stable.  
 Thus, the large black hole dominates at the unstable solutions at large baryochemical potentials. 
It is also interesting to note in passing that the crossing point of the two Gibbs free energies, around $\mu_B \approx 500$ MeV \cite{Fodor:2001pe,Ejiri:2003dc,Stephanov:2004wx}
 seems to coincide with the speculated position of the critical point of QCD.

The dual interpretation of 
$\Lambda$ remains to be clarified.  Here we present several possibilities.  (1) $\Lambda$ could be associated with the gluon pressure in hadrons in the context of the QCD bag model.  This would mean that $\Lambda$ is not associated with the gas of hadrons, but rather with each individual hadron.  (2) $\Lambda$ could be associated with a QCD chiral symmetry breaking scale.  (3)  $\Lambda$ could be  associated with a QCD confinement scale.  Further investigation will be required to see which of these provides the best dual description.

\section{Conclusions}\label{sec:conclusions}

We have shown that the conjectured equivalence between gravitational confinement and color confinement, implemented for asymptotically flat charged black holes,
can be fruitfully extended to include a negative cosmological constant. This quantity can be interpreted as a thermodynamic pressure $P$ on the gravitational side.  Constraints from the AdS geometry yield two possible solutions in a phenomenological interesting region.  
\\
One, corresponding to the small charged black hole solution, yields a good fit with existing data in
the $T$ vs. $\mu$ plane for $0.95 < P/P_E < 0.75$, where $P_E$ is the extremal pressure.  The instability of the QCD plasma is consistent with the thermodynamic instability of the small black hole. 

The next step of this analysis would be to study the critical phenomena that can happen in a hot dense systems of quarks and gluons.  Indeed in gauge/gravity duality the Hawking-Page transition has been associated with a confinement/deconfinement transition.

\section*{Acknowledgements}

This work was supported in part by the Natural Sciences and Engineering Research Council of Canada. This work has been performed in the framework of COST Action CA15213 THOR.

\medskip
 
\bibliography{QCDBH}

\end{document}